\documentclass[aps,prl,groupedaddress,showpacs,a4paper,twocolumn]{revtex4}

\usepackage{graphicx}                 
\usepackage{amsmath,amssymb,amsfonts} 

\DeclareFontFamily{OML}{eur}{\skewchar\font127}
\DeclareFontShape{OML}{eur}{m}{n}{<5> <6> <7> <8> <9> gen * eurm <10> <10.95>
  <12> <14.4> <17.28> <20.74> <24.88> eurm10}{}
\DeclareSymbolFont{greek}{OML}{eur}{m}{n}
\DeclareMathSymbol{\micro}{\mathord}{greek}{"16}

\begin{document}

\title{Light Filaments Without Self Guiding}

\author{%
  Audrius Dubietis,$^1$
  Eugenijus Gai\v{z}auskas,$^1$
  Gintaras Tamo\v{s}auskas,$^1$
  and
  Paolo Di Trapani$^2$
}

\affiliation{$^1$
  Department of Quantum Electronics, Vilnius University, Sauletekio al.~9,
  LT-2040 Vilnius, Lithuania
}

\affiliation{$^2$
  Istituto Nazionale di Fisica della Materia (INFM) and Department of Physics,
  University of Insubria, Via Valleggio~11, IT-22100 Como, Italy
}


\begin{abstract}
  An examination of the propagation of intense 200\,fs pulses in water
  reveals light filaments not sustained by the balance between Kerr-induced
  self-focusing and plasma-induced defocusing. Their appearance is interpreted
  as the consequence of a spontaneous reshaping of the wave packet
  form a gaussian into a conical wave, driven by
  the requirement of maximum localization, minimum losses and stationarity in
  the presence of non-linear absorption.
\end{abstract}


\maketitle

Since its discovery by Braun \emph{et al.}~\cite{Braun} in 1995,
the spontaneous filament formation accompanying intense
fs\nobreakdash-pulse propagation in air has received rapidly
increasing attention, both for the generation of coherent, soft
X\nobreakdash-rays~\cite{Lange98a}, IR~\cite{Kasparian}, as well
as sub-terahertz~\cite{Tzortzakis02} radiation, and for remote
sensing in the atmosphere~\cite{Kasparian03}. More recently
research has been extended to the case of filament formation in
condensed matter: namely, in fused silica~\cite{Tzortzakis01} and
water~\cite{Liu, Dubietis}, mainly in connection with fundamental
investigation in the soliton field~\cite{Henz, Berge}. In spite of
the different power and length scales that the process exhibits,
two key features emerge, which are substantially the same in all
media investigated: (i) for powers well exceeding the critical
value for continuous-wave (CW) self-focusing, a light filament
appears and propagates in the absence of diffraction or optical
breakdown for several to many diffraction lengths; and (ii) the
filament only contains a small fraction of the total beam power,
limited by the so-called ``intensity clamping
effect''~\cite{Liu02}, while the residual excess remains in the
filament periphery with no apparent trapping.
\begin{figure}[hbt]
  \includegraphics[width=9cm]{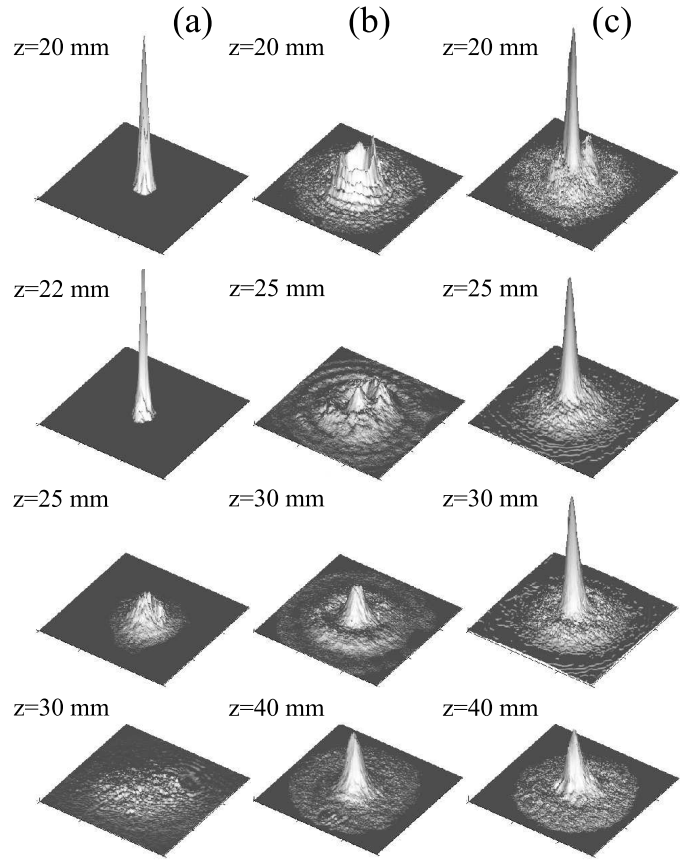}
  \caption{%
    Profiles of (a) clipped, (b) stopped and (c) free filaments for
  $20\,$mm$<z<40\,$mm. The input peak power is $P_p=13\,P_{cr}$.
    \label{fig:death}
  }
\end{figure}

Two major approaches have been proposed for the modeling,
description and understanding of the underlying physics: the first
interprets the filament as a genuine soliton-like, self-guided
beam, whose stationarity is supported by the dynamical balance
between Kerr-induced self-focusing and plasma-induced defocusing,
where plasma formation is due to multi-photon absorption
(MPA)~\cite{Berge00}. In this scenario the role of the non-trapped
radiation is marginal. In fact, although compensation of
MPA-induced power losses by external radiation has been addressed
experimentally~\cite{Nibbering}, this refilling effect was only
interpreted as perturbative, \emph{i.e.}, useful for prolonging
the filament life while not being structural to its existence. The
second approach, in contrast, only considers the filament as an
optical illusion related to the use of time-integrated detection.
The models, based on moving focus~\cite{Brodeur} or dynamic
spatial replenishment~\cite{Mlejnek}, interpret the filament as
continuously absorbed and regenerated by subsequent focusing of
different \emph{temporal} slices of the pulse.

In this letter we demonstrate how, for the case investigated of $200\,$fs
filaments in water, energy refilling by the surrounding radiation is so
important that one can by no means consider the filament in terms of a genuine,
self-guided beam. Our results indicate that filament regeneration can well be
described within the framework of a purely \emph{spatial} process, with the
fake self-guided beam as resulting from subsequent focusing of different
spatial coronas of the beam. The key picture we propose as a physical
interpretation of the entire dynamics addresses the spontaneous transformation
of a Gaussian into a Bessel-type beam, driven by the requirements of minimum
(non-linear) losses, maximum stationarity and maximum localization.

The criterion we adopted for distinguishing between a true (\emph{e.g.},
self-guided) filament and a fake (\emph{e.g.}, strongly refilled) is based on
the simple assumption that a genuine filament should survive (for at least 2--3
diffraction lengths) after transmission through a pinhole that removes the
non-trapped part of the radiation and transmits most of the filament energy
content. The Kerr medium that we have chosen is water; with respect to
solid-state media, it has the key advantages of being free from damage
problems, of allowing easy and continuous scanning of sample length, and of
permitting the insertion of pinholes along the beam path \emph{inside} the
non-linear media; in comparison with air, water has much less turbulence, which
results in higher beam-pointing stability and so allows the use of pinholes of
comparable size to the filament. We note that this was not the case for the
experiment in air quoted above~\cite{Nibbering}, where a very large
(\emph{e.g.}, $1\,$mm) pinhole was used.

The experiment was performed by launching a $527\,$nm,
$\mathop\sim3\,\micro$J, $200\,$fs, spatially filtered beam with
an approximately $0.1\,$mm FWHM waist located at the input facet
of a water-filled cuvette, and then monitoring the output-beam
fluence profile. The wave packet launched was provided by an
SHG-compressed, CPA Nd:glass laser (TWINKLE, Light Conversion
Ltd.), operated at a $33\,$Hz repetition rate. The cuvette was
made of $1\,$mm thick, syringe-shaped quartz, which allowed
continuous tuning of the sample length in the range
$z=5$--$40\,$mm. The output beam was imaged onto a CCD camera
(8-bit Pulnix TM-6CN) by an $f=+50\,$mm achromatic objective, with
$8\times$ magnification. The results highlighted the appearance of
a single filament with almost constant $\mathop\sim20\,\micro$m
FWHM diameter in the $z=15$--$40\,$mm range investigated.

Figure~\ref{fig:death}(a) shows the effect of inserting a
$55\,\micro$m-diameter pinhole in the water cuvette at $z=20\,$mm.
The energy transmitted by the pinhole was $20\%$ of the total
incident energy. Although the transmitted beam attempted to focus
after a short distance, (\emph{e.g.}, within one diffraction
length; see the plot at $z=22\,$mm), the filament survived no
further and we observed a rapid decay with divergence
$\mathop\sim2\times$ larger than that of a Gaussian beam of the
same FWHM diameter. It is necessary to double check whether, in
the absence of pinhole, the filament survives just because it is
strongly supported by energy refilling from the outside beam. To
do this, we also performed the complementary experiment, by
blocking the central spike with the sole aid of a $55\,\micro$m
beam stopper, printed on a $100\,\micro$m-thick BK7 glass plate
and inserted into the beam path at $z=20\,$mm. As depicted in
Fig.~\ref{fig:death}(b), a central spike of the original
dimensions reappeared at $z=25\,$mm, gaining power as it
propagated. The profiles measured in the case of ``free-filament''
propagation are reported in Fig.~\ref{fig:death}(c) for
comparison, in the same $z$ range. Note how the effect of the beam
stopper is barely detectable after only $20\,$mm of propagation.
These results unequivocally show that the observed filament can by
no means be described in terms of a soliton, beam self-guiding
effect.
\begin{figure}[hbt]
  \includegraphics[width=9cm]{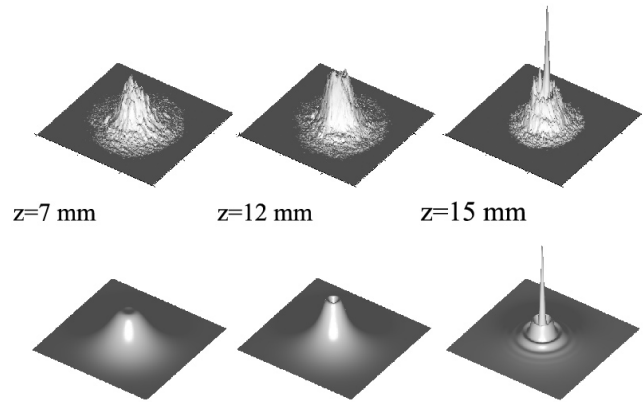}
  \caption{%
    Transient stage of the filament formation: (top) experimental,
    $P_{p}=17\,P_{cr}$; (bottom) numerical, $P=25\,P_{cr}$,
    $\beta^{(4)}\simeq2.0\times10^{-34}\,$cm$^5$/W$^3$.
    \label{fig:transient}
  }
\end{figure}

In what follows, we intend to demonstrate how the key features of
the filament-formation process may be interpreted as the effect of
spontaneous evolution of the input wave-packet toward a \emph{non
soliton-like}, but still \emph{stationary}, profile, an hypothesis
that has never before been considered. In order to make the point
clear, we propose here the \emph{simplest possible model} that
fully supports such a scenario,
by adopting the quite severe CW approximation, though \emph{a
priori} exclusion of a possible contribution of temporal or
spatio-temporal (ST) effects was not possible. In fact, the
results shown in Fig.~\ref{fig:death} are also compatible with
standard, moving-focus~\cite{Brodeur} or dynamic
spatial-replenishment~\cite{Mlejnek} models (this can be
understood by noting that if different \emph{temporal} slices are
focused at different planes, pinholes or stoppers could have
different effects on different temporal portions of the
wave-packet). Indeed, close inspection of ST wave-packet profiles
by means of a high-resolution ($20\,$fs) 3D-mapping technique (in
the case of input pulses shorter by a factor two) revealed a
dynamics that requires the full 3D model to be described
precisely~\cite{ST_NLGW}. However, such \emph{ST effects}, the
importance of which dramatically increases on shortening the
pulses, are \emph{not necessary for filament formation}. From the
experimental viewpoint, this claim is strongly supported by the
measurements with five times longer pulses (we do not present the
data here), whose dynamics appeared to be very similar to the
present case, especially in the asymptotic regime. Further support
comes \emph{a posteriori} from the rather good matching achieved
between experiment and CW calculations. As a second, highly
qualifying, approximation, we \emph{have neglected the role of the
plasma-defocusing effect}, thus avoiding the occurrence of any
self-guiding regime. In so doing, we wish to focus attention on
the final stage of the dynamics, where plasma-induced effects are
supposed to play only a minor role, seeking the occurrence of a
``final state'' that behaves as an ``attractor'' for the entire
dynamics. The only non-linear terms included in the model are,
therefore, those of self-focusing and MPA. Within the framework of
the paraxial approximation adopted, the resulting CW, elliptical,
modified non-linear Schr\"{o}dinger equation for the field
amplitude $A$ reads:
\begin{align}
  \frac{\partial A}{\partial z} &=
  \frac{i}{2k}
  \left(
    \frac{\partial^2\!A}{\partial x^2} +
    \frac{\partial^2\!A}{\partial y^2}
  \right)
  \nonumber
\\
  & \quad \null
  + \frac{i\omega_0 n_2}{c} |A|^2\!A
  - \frac{\beta^{(K)}}{2} |A|^{2K-2} A ,
  \label{eq:NLS}
\end{align}
where $z$ is propagation distance,
$n_2=2.7\times10^{-16}\,$cm$^2/$W is the non-linear refractive
index, and $\beta^{(K)}$ is the MPA coefficient. In our model we
took $K=4$ in order to account for three-photon absorbtion (water
band-gap $W_g=6.5\,$eV, photon energy $2.4\,$eV) and, to a first
approximation, for further absorption due to the photo-induced
plasma. We verified that similar results are also obtained with
different $K$. Owing to the over-simplification of our model,
which does not allows for \emph{ab initio} evaluation of
parameters, for the absorption coefficient $\beta^{(4)}$ we took
the value that better fitted the experimental results (see the
figure caption). The fit led to a slight dependence of
$\beta^{(4)}$ on input beam power, which may be understood as
owing to the different role played by plasma absorption. We
consider evolution of a linearly polarized beam propagating with
central frequency $\omega_0$ and wave-number $k=\omega_0n/{c}$,
with $n=1.334$. The resulting critical power for CW beam collapse
is $P_{cr}=3.77{\lambda^2}/(8{\pi}nn_2)=1.15\,$MW. The input beam
profile was taken as Gaussian with a $0.12\,$mm diameter at FWHM.
Note the slightly larger diameter (and also larger powers, see the
figure captions) that had to be taken with respect to the
experiment, in order to compensate for a slight overestimation of
the losses by our model. Equation~\eqref{eq:NLS} was solved
numerically by means of the FFT Runge-Kutta split-step method.

A very surprising result is that our model, designed for a
qualitative description of the asymptotics, also turned out to
describe the very transient regime rather well, where plasma is
eventually supposed to play its major role. Both the experimental
and numerical results shown in Fig.~\ref{fig:transient} indicate
the occurrence of an appreciable flattening of the top-beam
profile in the first $7\,$mm of propagation (left column), which
occurs in conjunction with a slight beam self-focusing. Owing to
the flattop profile, the major effect of this self-focusing
becomes that of developing a sharp and rather intense ring peak on
the external border of the ``mesa'' (see the ``volcano'' profile
shown in the central column of the figure). Owing to the Kerr
non-linear response, this ring-like modulation gives rise to a
short, positive, toroidal lens, which focuses the outside corona
(the ``skin'') of the beam and leads the very intense spike to
appear in the center (right column in the figure). Owing to its
very high intensity, this spike leads to a very strong, though
only local, MPA. The persisting toroidal lens, however, keeps
refilling the spike, thus establishing a quasi stationary regime
(filament), which remains as long as most of the beam power is
coupled to the central spike.

\begin{figure}[hbt]
  \includegraphics[width=9cm]{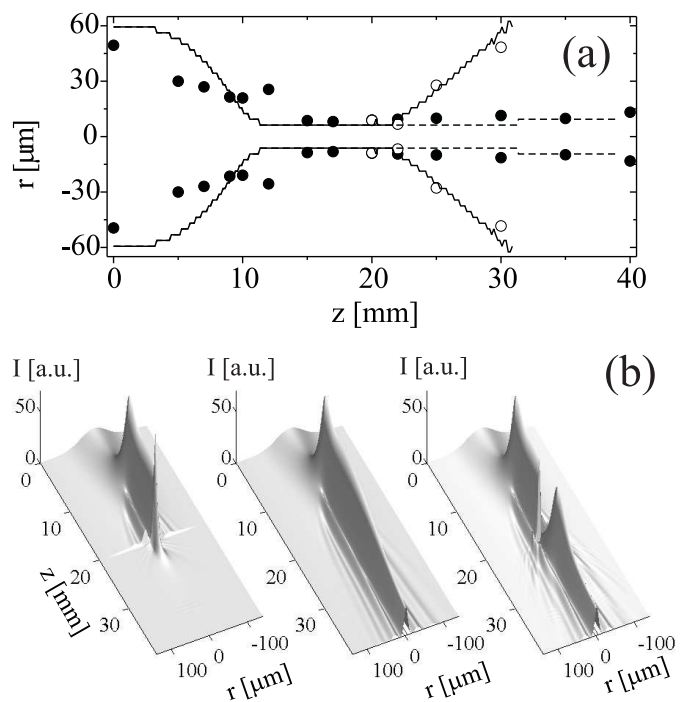}
  \caption{%
  (a) HWHM beam radius \emph{vs.}\ $z$: experiment (full circles) and
  simulation (dashed line). Open circles and solid line: the same in the case
  of a $55\,\micro$m pinhole only transmitting the central spike.
  (b) Calculated transverse intensity profiles for the clipped (left),
  free (center) and stopped (right) filament case. Experiment: $P_{p}=13\,P_{cr}$.
  Numerical values: $P=15\,P_{cr}$;
  $\beta^{(4)}\simeq1.25\times10^{-34}\,$cm$^5$/W$^3$.
  \label{fig:exp&model}
  }
\end{figure}

Figure~\ref{fig:exp&model}(a) presents a comprehensive summary of the measured
and calculated dynamics, by plotting the FWHM beam diameters in the entire
(\emph{i.e.}, transient plus quasi-stationary) range investigated.
Both the free-filament and the $55\,\micro$m clipped filament are
considered. Note the rather impressive capability of our model in
describing the apparent stationary-filament regime, \emph{in the
absence of any defocusing effect}. Note also how the model
predicts quenching of the filament, when the pinhole is inserted.
For a more complete description of the dynamics, the corresponding
calculated transverse intensity profiles \emph{vs.}\ $z$ are
reported in Fig.~\ref{fig:exp&model}(b). Note that here the case
of the $55\,\micro$m beam stopper is also depicted (right).

\begin{figure}[hbt]
  \includegraphics[width=9cm]{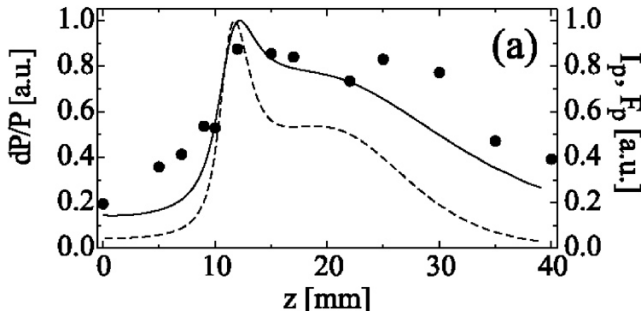}
 \caption{%
  Calculated peak intensity $I_p$ (solid line); measured peak fluences $F_p$
  (full circles); calculated fractional power losses (dashed curve), for the
  same conditions as in Fig.~\ref{fig:exp&model}.
  \label{fig:losses}
  }
\end{figure}

In order to elaborate the physical interpretation of the entire
dynamics, in Fig.~\ref{fig:losses}(a) we plot the (measured and
calculated) peak fluence $F_p$ and intensity $I_p$, and the
(calculated only) fractional power losses $dP/P$ \emph{vs.}\ $z$.
The results clearly indicate the occurrence of two different
regimes: for $z<12\,$mm, both $I_p$ ($F_p$) and $dP/P$ increase
due to the overall effect of self-focusing and of MPA. After that,
however, while $I_p$ ($F_p$) keeps a high and almost constant
value, a sharp drop of $dP/P$ occurs. Close inspection of the
results in Fig.~\ref{fig:exp&model}(b) indicates that the beam has
undergone an important transformation: from a Gaussian to
Bessel-like profile, which has preserved appreciable localization
(\emph{e.g.}, high $I_p$) and stationarity while minimizing the
non-linear losses (MPA). In fact, owing to the presence of "cold",
slowly decaying tails, which contain most of beam power, a conical
wave is provided with a large reservoir that can refuel the "hot"
central spike and so preserve stationarity, in spite of the
presence of non-linear losses. If the robustness of the Bessel
profile under Kerr-induced spatial phase modulation is accounted
for~\cite{Bessel}, one could foresee that there might indeed exist
a Bessel-like, infinite-power, exact stationary solution of
Eq.~\eqref{eq:NLS} that behaves as strong attractor for the entire
non-linear beam transformation.

In conclusion, by clipping or stopping a light filament while it
propagates in water we have shown experimentally that it does not
behave as a self-guided wave packet, being structurally sustained
by a strong energy flux from the surrounding beam. The matching
obtained between the experimental results and those based on a CW
model that only accounts for self-focusing and non-linear losses
provides a strong indication that all the temporal effects as well
as all those related to plasma-induced defocusing are not
essential to the occurrence of the apparent guiding effect. In
contrast, the quasi-stationary filament appears as the dynamical
balance between sharp focusing of the very external periphery of
the beam (skin focusing effect) and non-linear losses, which
continue to absorb the generated spike. We interpret the entire
dynamics as the attempt of the wave-packet to readjust its shape,
due to the non-linear coupling, according to the requirements of
maximum localization, minimum losses and stationarity. We predict
the existence of an exact, infinite-power, Bessel-like stationary
solution of the modified CW non-linear Schr\"{o}dinger equation in
the presence of non-linear losses.

\begin{acknowledgments}
  The authors acknowledge assistance in numerical calculations by A.~Parola,
  M.~Porras and A.~Berzanskis, discussions with R.~Danielius and A.~Piskarskas,
  technical support in the measurements by E.~Ku\u{c}inskas and financial
  support by MIUR (Cofin01/FIRB01) and EC~CEBIOLA (ICA1-CT-2000-70027)
  contracts.
\end{acknowledgments}

\end{document}